\documentclass[fp,twocolumn]{jpsj3}
\usepackage{graphicx}
\usepackage{color}
\usepackage{bm}

\title{%
Electron Transport along Screw Dislocations in a Strong Topological Insulator
}

\author{%
Tatsuro Sakaguchi and Yositake Takane
}

\inst{%
Graduate School of Advanced Science and Engineering,\\
Hiroshima University, Higashihiroshima, Hiroshima 739-8530, Japan
}

\recdate{ \hspace{50mm} }

\abst{%
In a three-dimensional strong topological insulator,
gapless helical surface states appear everywhere on its surface.
In the presence of a screw dislocation,
gapless helical modes also appear in the vicinity of
the corresponding dislocation line.
Let us focus on a case where a pair of screw dislocations connects
the top and bottom surfaces of a strong topological insulator
with a shape of rectangular parallelepiped.
The dislocation-induced helical modes are expected to act as
one-dimensional conduction channels connecting the top and bottom surfaces.
To examine this expectation, we calculate the two-terminal conductance
between a pair of electrodes placed on the top and bottom surfaces.
We found that the dislocation-induced helical modes
and the helical surface states on the side surface
contribute to the two-terminal conductance.
The contribution of the dislocation-induced helical modes
becomes more dominant than
that of the helical surface states in certain situations.
}

\begin{document}
\maketitle

\section{Introduction}

Three-dimensional $\mathbb{Z}_{2}$ topological insulators
are characterized by strong and weak indices
$(\nu_{0}, \nu_{x}\nu_{y}\nu_{z})$.~\cite{fu,moore,roy,hasan}
A strong topological insulator with $\nu_{0}=1$ hosts gapless topological
boundary states everywhere on its surface.
These two-dimensional states,
which are protected by time-reversal symmetry,
are helical because they exhibit helical spin texture,
i.e., the spin orientation of a surface state is tied to its momentum.
This helical nature of gapless surface states also manifests itself
in topological states induced by lattice defects.
When a screw or edge dislocation is inserted into
a strong topological insulator,
a pair of counterpropagating gapless modes appears in the vicinity of
the corresponding dislocation line under a certain condition
[see Eq.~(\ref{eq:condition-helical})].~\cite{ran,zhang}
These one-dimensional modes are also helical in that
the spin orientation of a one-dimensional state determines
its propagating direction, i.e.,
if the gapless mode propagating in one direction consists of spin-up states,
that propagating in the other direction consists of spin-down states.
The appearance of one-dimensional gapless modes due to the insertion of
a screw or edge dislocation has also been demonstrated not only in
gapped topological systems~\cite{teo,imura1,yoshimura,shiozaki,slager,pauly}
but also in Weyl and Dirac semimetals~\cite{imura2,sumiyoshi,chernodub,
kodama,huang,soto-garrido,amitani,zheng}.

For definiteness, let us consider a $\mathbb{Z}_{2}$ topological insulator of
rectangular parallelepiped shape into which
a pair of screw dislocations is inserted [see Fig.~1(a)].
We assume that the screw dislocations are parallel to the $z$-axis
and connect the top and bottom surfaces of the system.
A step edge is induced on the top and bottom surfaces
because of the presence of the screw dislocations [see Fig.~1(b)].
In the weak topological insulator case of $\nu_{0} = 0$,
we can set up the system such that a pair of
counterpropagating helical modes appears along the loop consisting of
two dislocation lines and two step edges.~\cite{yoshimura}
In this setup, helical surface states appear only on the side surface
without penetrating the top and bottom surfaces.
Therefore, the one-dimensional helical modes are isolated
from the helical surface states.
In the strong topological insulator case of $\nu_{0} = 1$,
helical surface states appear on all surfaces.
Hence, the one-dimensional helical modes are hybridized with
the helical surface states on the top and bottom surfaces near the step edges.
In the presence of such hybridization, we expect that the one-dimensional
helical modes along each dislocation line act as conduction channels
connecting the top and bottom surfaces.~\cite{hamasaki1,hamasaki2}

We examine whether the dislocation-induced helical modes
act as conduction channels for electron transport.
A simple way to do this is to calculate
the two-terminal conductance $\mathcal{G}$ between two electrodes
one of which is placed on the top surface and
the other one is placed on the bottom surface.
The dislocation-induced helical modes hybridized with
the helical surface states contribute to $\mathcal{G}$.
However, they are not the only channels connecting the two electrodes.
Because the helical surface states themselves directly connect
the electrodes through the side surface,
they also contribute to $\mathcal{G}$.
In contrast, the contribution of bulk states is negligible
if the Fermi energy is in the subgap region of
the strong topological insulator.
Thus, $\mathcal{G}$ is governed by the dislocation-induced helical modes
and the helical surface states.
We must separate these two contributions to examine the role of
the dislocation-induced helical modes.

In this paper, we calculate the two-terminal conductance $\mathcal{G}$
between a pair of electrodes placed on the top and bottom surfaces of
a strong topological insulator with a shape of rectangular parallelepiped.
We found that the dislocation-induced helical modes dominantly contribute to
$\mathcal{G}$
when the helical surface states on the side surface are eliminated
by imposing a periodic boundary condition in the $x$- and $y$-directions.
If the helical surface states on the side surface are considered
under an open boundary condition in the $x$- and $y$-directions,
the contribution of the dislocation-induced helical modes does not
dominate that of the helical surface states.
However, the former contribution dominates the latter contribution
if the area of the top and bottom surfaces is increased
while keeping the areal density of the dislocation pairs constant.
We also found that the dislocation-induced helical modes
become dominant if an energy gap is opened in the spectrum of the
helical surface states on the side surface
by applying a local perturbation that breaks the time-reversal symmetry.

In the next section, we introduce a tight-binding Hamiltonian
for a $\mathbb{Z}_{2}$ topological insulator.
In Sect.~3, we briefly describe the method for calculating
the two-terminal conductance $\mathcal{G}$ between two electrodes.
In Sect.~4, we provide the numerical results of $\mathcal{G}$ for two cases:
a case without the side surface
under a periodic boundary condition in the $x$- and $y$-directions,
and a case with the side surface
under an open boundary condition in the $x$- and $y$-directions.
The last section is devoted to a summary and short discussion.

\section{Model}

We introduce a tight-binding model for a $\mathbb{Z}_{2}$ topological insulator
on a cubic lattice with a lattice constant $a$
in a rectangular parallelepiped shape of volume
$N_{x} \times N_{y} \times N_{z}$.
The indices $l$, $m$, and $n$ are used to specify lattice sites
in the $x$-, $y$-, and $z$-directions, respectively, where
$1 \le l \le N_{x}$, $1 \le m \le N_{y}$, and $1 \le n \le N_{z}$.
The four-component state vector for the $(l,m,n)$th site is expressed as
\begin{align}
  |l,m,n \rangle
    = \Bigl[ |l,m,n \rangle_{1\uparrow} \hspace{0.8mm}
             |l,m,n \rangle_{2\uparrow} \hspace{0.8mm}
             |l,m,n \rangle_{1\downarrow} \hspace{0.8mm}
             |l,m,n \rangle_{2\downarrow}
      \Bigr] ,
\end{align}
where each lattice site has orbital and spin degrees of freedom described by
$\eta = 1, 2$ and $\sigma = \uparrow, \downarrow$, respectively.
The Hamiltonian is given by $H = H_{\rm d}+H_{x}+H_{y}+H_{z}$
with~\cite{liu}
\begin{align}
    \label{eq:H_d}
   H_{\rm d}
 & = \sum_{l=1}^{N_{x}}\sum_{m=1}^{N_{y}}\sum_{n=1}^{N_{z}}
     |l,m,n \rangle \mathcal{H}_{d} \langle l,m,n| ,
         \\
    \label{eq:H_x}
   H_{x}
 & = \sum_{l=1}^{N_{x}-1}\sum_{m=1}^{N_{y}}\sum_{n=1}^{N_{z}}
     |l+1,m,n \rangle \mathcal{H}_{x} \langle l,m,n|
     + {\rm h.c.} ,
        \\
    \label{eq:H_y}
   H_{y}
 & = \sum_{l=1}^{N_{x}}\sum_{m=1}^{N_{y}-1}\sum_{n=1}^{N_{z}}
     |l,m+1,n \rangle \mathcal{H}_{y} \langle l,m,n|
     + {\rm h.c.} ,
        \\
    \label{eq:H_z}
   H_{z}
 & = \sum_{l=1}^{N_{x}}\sum_{m=1}^{N_{y}}\sum_{n=1}^{N_{z}-1}
     |l,m,n+1 \rangle \mathcal{H}_{z} \langle l,m,n|
     + {\rm h.c.} ,
\end{align}
where
\begin{align}
   \mathcal{H}_{d}
 & = \left[ 
       \begin{array}{cc}
         M\tau_{z} &
         0_{2 \times 2} \\
         0_{2 \times 2} &
         M\tau_{z}
       \end{array}
     \right] ,
       \\
   \mathcal{H}_{x}
 & = \left[ 
       \begin{array}{cc}
         -m_{\parallel}\tau_{z} &
         \frac{i}{2}A\tau_{x} \\
         \frac{i}{2}A\tau_{x} &
         -m_{\parallel}\tau_{z}
       \end{array}
     \right] ,
       \\
   \mathcal{H}_{y}
 & = \left[ 
       \begin{array}{cc}
         -m_{\parallel}\tau_{z} &
         \frac{1}{2}A\tau_{x} \\
         -\frac{1}{2}A\tau_{x} &
         -m_{\parallel}\tau_{z}
       \end{array}
     \right] ,
       \\
   \mathcal{H}_{z}
 & = \left[ 
       \begin{array}{cc}
         -m_{\perp}\tau_{z} + \frac{i}{2}B\tau_{x}&
         0_{2 \times 2} \\
         0_{2 \times 2} &
         -m_{\perp}\tau_{z} - \frac{i}{2}B\tau_{x}
       \end{array}
     \right] .
\end{align}
Here, $M$ is given by
\begin{align}
  M = m_{0} + 4m_{\parallel} + 2m_{\perp} ,
\end{align}
and $\tau_{q}$ is the $q$-component of Pauli matrices ($q \in x,y$, and $z$).
The strong and weak indices $(\nu_{0}, \nu_{x}\nu_{y}\nu_{z})$ characterizing
a $\mathbb{Z}_{2}$ topological insulator are determined by the parameters
$m_{0}$, $m_{\parallel}$, and $m_{\perp}$.
We set $m_{\perp}/m_{\parallel} = 1$ and $m_{0}/m_{\parallel} = - 10$
throughout this paper.
With this setting, the Hamiltonian $H$ describes
the strong topological insulator with
$(\nu_{0}, \nu_{x}\nu_{y}\nu_{z}) = (1,111)$.~\cite{imura3}

The Hamiltonian, consisting of the four terms specified in
Eqs.~(\ref{eq:H_d})--(\ref{eq:H_z}), describes a system
under an open boundary condition in the $x$-, $y$-, and $z$-directions.
In addition to the system under the open boundary condition
in the three directions,
a system under a periodic boundary condition
in the $x$- and $y$-directions (see Sect.~4) is also considered
to eliminate the helical surface states on the side surface.
The periodic boundary condition in the two directions is implemented
if the following boundary terms are added to $H$:
\begin{align}
   \Delta H_{x}
 & = \sum_{m=1}^{N_{y}}\sum_{n=1}^{N_{z}}
     |1,m,n \rangle \mathcal{H}_{x} \langle N_{x},m,n|
     + {\rm h.c.} ,
        \\
   \Delta H_{y}
 & = \sum_{l=1}^{N_{x}}\sum_{n=1}^{N_{z}}
     |l,1,n \rangle \mathcal{H}_{y} \langle l,N_{y},n|
     + {\rm h.c.} 
\end{align}

The bulk spectrum of this model is given as
$E(\mib{k}) = \pm \sqrt{\Xi({\mib k})}$ in the infinite volume limit,
where ${\mib k} = (k_{x}, k_{y}, k_{z})$ is a wave vector and
\begin{align}
 \Xi({\mib k})
     & = [M - 2m_{\parallel}\left(\cos(k_{x}a) + \cos(k_{y}a)\right)
            - 2m_{\perp}\cos(k_{z}a)]^{2}
     \nonumber \\
     & \hspace{-2mm}
        + A^{2}\left(\sin^{2}(k_{x}a) + \sin^{2}(k_{y}a)\right)
        + B^{2}\sin^{2}(k_{z}a) .
\end{align}
In the case
where $m_{\perp}/m_{\parallel} = 1$ and $m_{0}/m_{\parallel} = - 10$,
the energy gap defined as $E_{\rm g} = \min_{\mib k}\{\sqrt{\Xi({\mib k})}\}$
is given by $E_{\rm g}/m_{\parallel} = 2$.

Then, let us introduce a pair of screw dislocations parallel to the $z$-axis
in our tight-binding model without deforming the lattice structure itself.
We assume that the screw dislocations are centered at
$(x_{\rm 1D},y_{\rm D}) = a(l_{\rm 1D}+ \frac{1}{2}, m_{\rm D}+ \frac{1}{2})$
and
$(x_{\rm 2D},y_{\rm D}) = a(l_{\rm 2D}+ \frac{1}{2}, m_{\rm D}+ \frac{1}{2})$
with
\begin{align}
  l_{\rm 1D} = \frac{N_{x}}{4} , \hspace{5mm}
  l_{\rm 2D} = \frac{3N_{x}}{4} , \hspace{5mm}
  m_{\rm D}  = \frac{N_{y}}{2} ,
\end{align}
where the origin of the $x$- and $y$-coordinates $(x,y)=(0,0)$
is set at $(l,m) = (0,0)$.
Here, it is implicitly assumed that $N_{x}$ and $N_{y}$ are integers
that are multiple of $4$ and $2$, respectively.
We also assume that the screw dislocations at
$(x_{\rm 1D},y_{\rm D})$ and $(x_{\rm 2D}, y_{\rm D})$
have a displacement of $N$ unit atomic layers,
and they are characterized by the Burgers vectors $\mib{b}_{1} = (0,0,N)$
and $\mib{b}_{2} = (0,0,-N)$, respectively.
Let us consider a slip plane with its two edges identical to
dislocation lines [see Fig.~1(a)]
and modify the hopping terms in $H$ across it so that each term connects
two different layers in the $z$-direction [see Fig.~1(b)].~\cite{imura1,imura2}
To do this, we reconnect the hopping terms in $H_{y}$ in the region of
$l_{\rm 1D} + 1 \le l \le l_{\rm 2D}$ by performing the following replacement:
\begin{align}
     \label{eq:reconnect1}
 & |l,m_{\rm D}+1,n \rangle \mathcal{H}_{y} \langle l,m_{\rm D},n|
             + {\rm h.c.}
       \nonumber \\
 & \to   
   |l,m_{\rm D}+1,n+N \rangle \mathcal{H}_{y} \langle l,m_{\rm D},n|
             + {\rm h.c.} ,
\end{align}
which shows that the site with $m = m_{\rm D}$ on the $n$th layer
is connected to the site with $m = m_{\rm D}+1$ on the $n+N$th layer
across the slip plane.
Consequently, $H_{y}$ becomes
\begin{align}
      H_{y}
 &  = \sum_{l=1}^{N_{x}}
      \sum_{\substack{m=1 \\ (m \neq m_{\rm D})}}^{N_{y}-1}
      \sum_{n=1}^{N_{z}}
      |l,m+1,n \rangle \mathcal{H}_{y} \langle l,m,n|
      + {\rm h.c.}
      \nonumber \\
 &  + \sum_{l=1}^{l_{\rm 1D}}\sum_{n=1}^{N_{z}}
      |l,m_{\rm D}+1,n \rangle \mathcal{H}_{y} \langle l,m_{\rm D},n|
      + {\rm h.c.}
      \nonumber \\
 &  + \sum_{l=l_{\rm 1D}+1}^{l_{\rm 2D}}\sum_{n=1}^{N_{z}-N}
      |l,m_{\rm D}+1,n+N \rangle \mathcal{H}_{y} \langle l,m_{\rm D},n|
      + {\rm h.c.}
      \nonumber \\
 &  + \sum_{l=l_{\rm 2D}+1}^{N_{x}}\sum_{n=1}^{N_{z}}
      |l,m_{\rm D}+1,n \rangle \mathcal{H}_{y} \langle l,m_{\rm D},n|
      + {\rm h.c.}
\end{align}
after the replacement.
A step edge with a length of $a(l_{\rm 2D}-l_{\rm 1D})$ is induced
by the screw dislocations on the top and bottom surfaces [see Fig.~1(b)].
\begin{figure}[btp]
\begin{center}
\includegraphics[height=6.0cm]{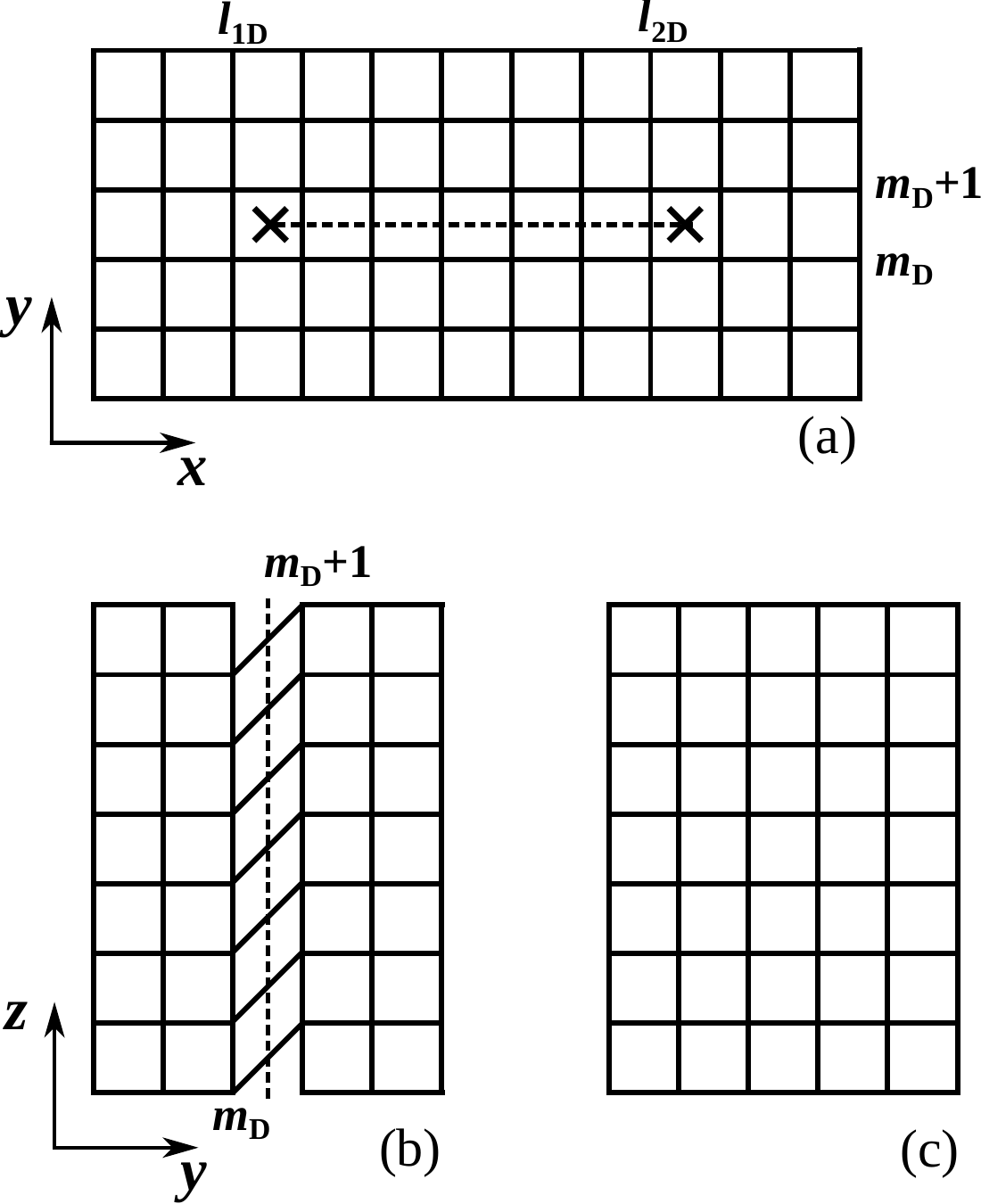}
\end{center}
\caption{
Lattice system with a pair of screw dislocations with
a displacement of one unit atomic layer (i.e., $N = 1$),
where each solid line between two neighboring sites represents
a hopping term in $H_{x}$, $H_{y}$, or $H_{z}$
that directly connects the two sites.
The dotted lines represent
the slip plane across which the hopping terms in $H$ are modified.
(a) Top view with the crosses representing the dislocation centers. 
(b) Side view of the modified region of $l_{\rm 1D} + 1 \le l \le l_{\rm 2D}$.
A step edge of one unit atomic layer is induced
by the screw dislocations at $m = m_{\rm D}$ on the top surface
and at $m = m_{\rm D}+1$ on the bottom surface.
(c) Side view of the unmodified region of
$l \le l_{\rm 1D}$ or $l_{\rm 2D} + 1 \le l$.
}
\end{figure}

As shown in Ref.~\citen{ran}, the appearance of one-dimensional gapless
helical modes confined in the vicinity of a screw dislocation
is determined by the Burgers vector $\mib{b}$ characterizing
the screw dislocation and the vector $\mib{M}$ defined in terms of
the weak indices as $\mib{M} = (\nu_{x},\nu_{y},\nu_{z})$.
The gapless helical modes appear if and only if
\begin{align}
     \label{eq:condition-helical}
  \mib{b} \cdot \mib{M} = 1   \hspace{8mm} (\bmod 2) .
\end{align}
In our setting, $\mib{M} = (1,1,1)$ and $\mib{b}_{1} = -\mib{b}_{2} = (0,0,N)$;
thus, the helical modes appear in our system when $N$ is an odd integer.

\section{Simulation of Electron Transport}

We study electron transport at zero temperature
between the top and bottom surfaces of a strong topological insulator
in the presence or absence of
a pair of screw dislocations parallel to the $z$-axis.
To characterize electron transport, we calculate
the two-terminal conductance between two electrodes,
where the first electrode is placed on the top surface
and the second one is placed on the bottom surface.
For simplicity, we assume that each electrode is coupled with
the top or bottom surface of the strong topological insulator
in a rectangular region
with $5$ sites in the $x$-direction and $4$ sites in the $y$-direction.
Hereafter, we use $\mib{r} \equiv (l,m,n)$ to denote the position
of each site in the system.
The rectangular region on the top surface and that on the bottom surface are
specified by $\mib{r}_{1} = (l, m, N_{z})$ and $\mib{r}_{2} = (l, m, 1)$,
respectively, with
$l_{\rm C}-2 \le l \le l_{\rm C}+2$ and $m_{\rm C}-1 \le m \le m_{\rm C}+2$.

We calculate the two-terminal conductance between two electrodes
using a formula in terms of Green's function.~\cite{meir,datta}
Let us define Green's function as
\begin{align}
  G = \left( E_{\rm F}{\mib 1}-H-\Sigma \right)^{-1} ,
\end{align}
where ${\mib 1}=\sum_{\mib r}|\mib{r}\rangle\langle\mib{r}|$,
$E_{\rm F}$ is the Fermi energy, and $\Sigma$ is the self-energy
describing the coupling of the system with the electrodes.
Here, $|\mib{r}\rangle$ represents $|l,m,n\rangle$.
We assume that $\Sigma = \Sigma_{1} + \Sigma_{2}$, with
\begin{align}
  \Sigma_{1}
  & =  -i\gamma \sum_{\mib{r}\in {\rm S}_{1}}|\mib{r}\rangle\langle\mib{r}| ,
  \\
  \Sigma_{2}
  & =  -i\gamma \sum_{\mib{r}\in {\rm S}_{2}}|\mib{r}\rangle\langle\mib{r}| ,
\end{align}
where $\gamma$ is the coupling strength and ${\rm S}_{p}$ denotes the set of
$5 \times 4$ sites in contact with the $p$th electrode ($p = 1, 2$).
In terms of Green's function, the transmission function
from the bottom electrode to the top electrode is defined as~\cite{meir}
\begin{align}
  T = \Tr \left\{\Gamma_{1} G \Gamma_{2} G^{\dagger}\right\} ,
\end{align}
where $\Gamma_{p} \equiv i(\Sigma_{p} -\Sigma_{p}^{\dagger})$.
The two-terminal conductance $\mathcal{G}$
between the top and bottom electrodes is expressed as
\begin{align}
  \mathcal{G} = \frac{e^2}{h}T .
\end{align}

In addition to $m_{\perp}/m_{\parallel} = 1$ and $m_{0}/m_{\parallel} = - 10$,
the following parameters are used: $A/m_{\parallel} = 1.5$,
$B/m_{\parallel} = 2.0$, and $\gamma/m_{\parallel} = 0.2$.

\section{Numerical Results}

In this section, we consider the transmission function $T$
instead of the two-terminal conductance $\mathcal{G}$.
We numerically calculate $T$ in the presence and absence of
the screw dislocations
in the cases with and without the side surface.
The system size is fixed as $N_{x} = 80$, $N_{y} = 40$, and $N_{z} = 52$.
The locations of the two electrodes are specified by
$l_{\rm C} = 20$ and $m_{\rm C} = 10$.
The screw dislocations are assumed to have a displacement of
one unit atomic layer (i.e., $N = 1$) unless otherwise noted.

\begin{figure}[btp]
\begin{center}
\includegraphics[height=4.0cm]{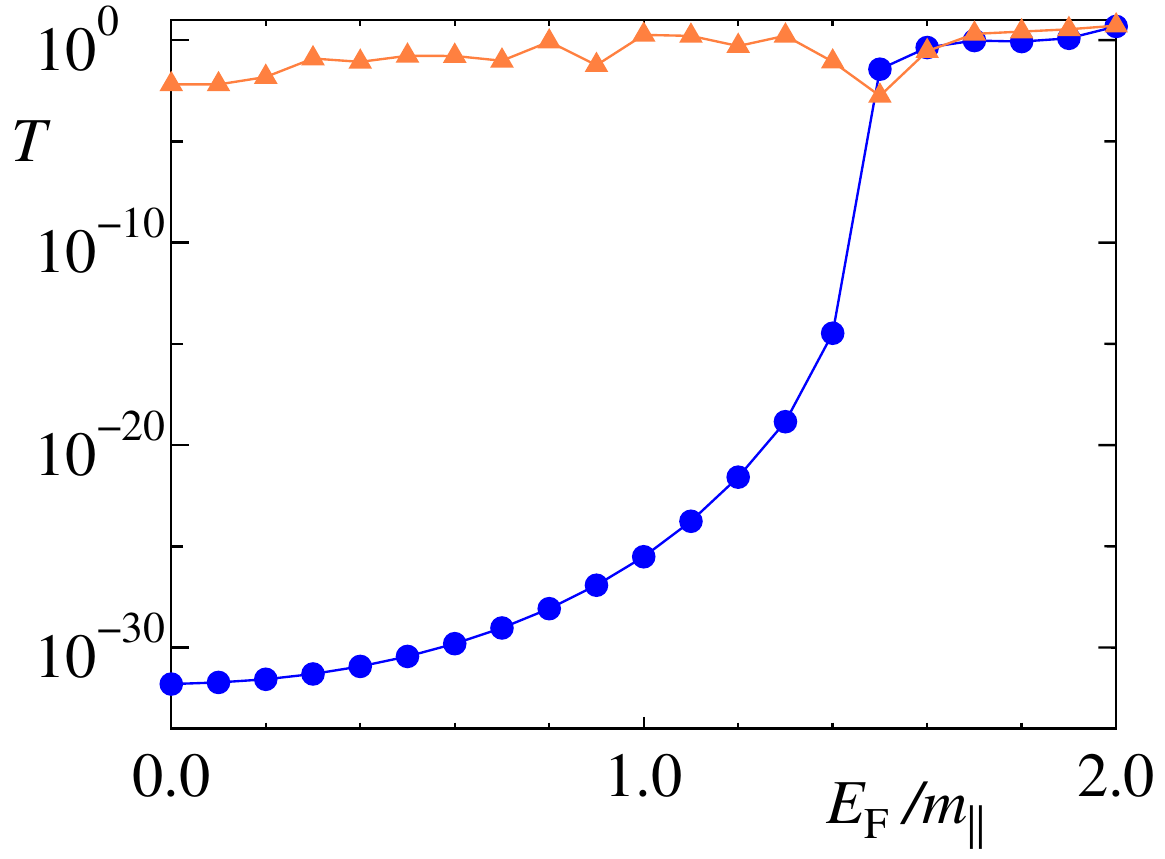}
\end{center}
\caption{
(Color online)
Numerical results of $T$ in the presence of the screw dislocations with $N = 1$
(triangles) and in the absence of the screw dislocations (circles)
for several values of $E_{\rm F}/m_{\parallel}$
under the periodic boundary condition in the $x$- and $y$-directions.
The solid lines serve as visual guides.
}
\end{figure}
Let us first consider the case without the side surface
under the periodic boundary condition in the $x$- and $y$-directions.
Figure~2 shows the numerical results of $T$ in the presence and absence of
the screw dislocations for various values of $E_{\rm F}$
in the range of $0 \le E_{\rm F}/m_{\parallel} \le 2$.
We observe that $T$ in the presence of the screw dislocations is
orders of magnitude larger than that in the absence of the screw dislocations
when $E_{\rm F}/m_{\parallel}$ is small,
showing that the dislocation-induced helical modes
certainly contribute to $T$.
The quantitative difference between the two cases becomes small
as $E_{\rm F}/m_{\parallel}$ increases
and almost vanishes when $E_{\rm F}/m_{\parallel} \gtrsim 1.6$.
This behavior is ascribed to the fact that the contribution of
the bulk states to $T$, which is almost independent
of the presence or absence of the screw dislocations,
increases rapidly as $E_{\rm F}$ approaches the gap edge.
Figure~2 indicates that
$E_{\rm F}/m_{\parallel} = 1.6$ corresponds to the gap edge.~\cite{comment1}

Recall that our setting of $m_{\perp}/m_{\parallel} = 1$ and
$m_{0}/m_{\parallel} = - 10$ results in
$(\nu_{0}, \nu_{x}\nu_{y}\nu_{z}) = (1,111)$.
As noted in Sect.~2, the dislocation-induced helical modes appear
in our system when $N$ is an odd integer.
To observe how $N$ affects $T$, we numerically calculate $T$ in the presence
of the screw dislocations at $E_{\rm F}/m_{\parallel} = 0.0$
for $N = 0$, $1$, $2$, $3$, $4$, and $5$,
where $N = 0$ corresponds to the case in the absence of the screw dislocations.
The numerical results shown in Fig.~3 indicate that
$T$ using an odd $N$ is orders of magnitude larger than that using an even $N$.
This confirms the theoretical expectation.
\begin{figure}[btp]
\begin{center}
\includegraphics[height=3.6cm]{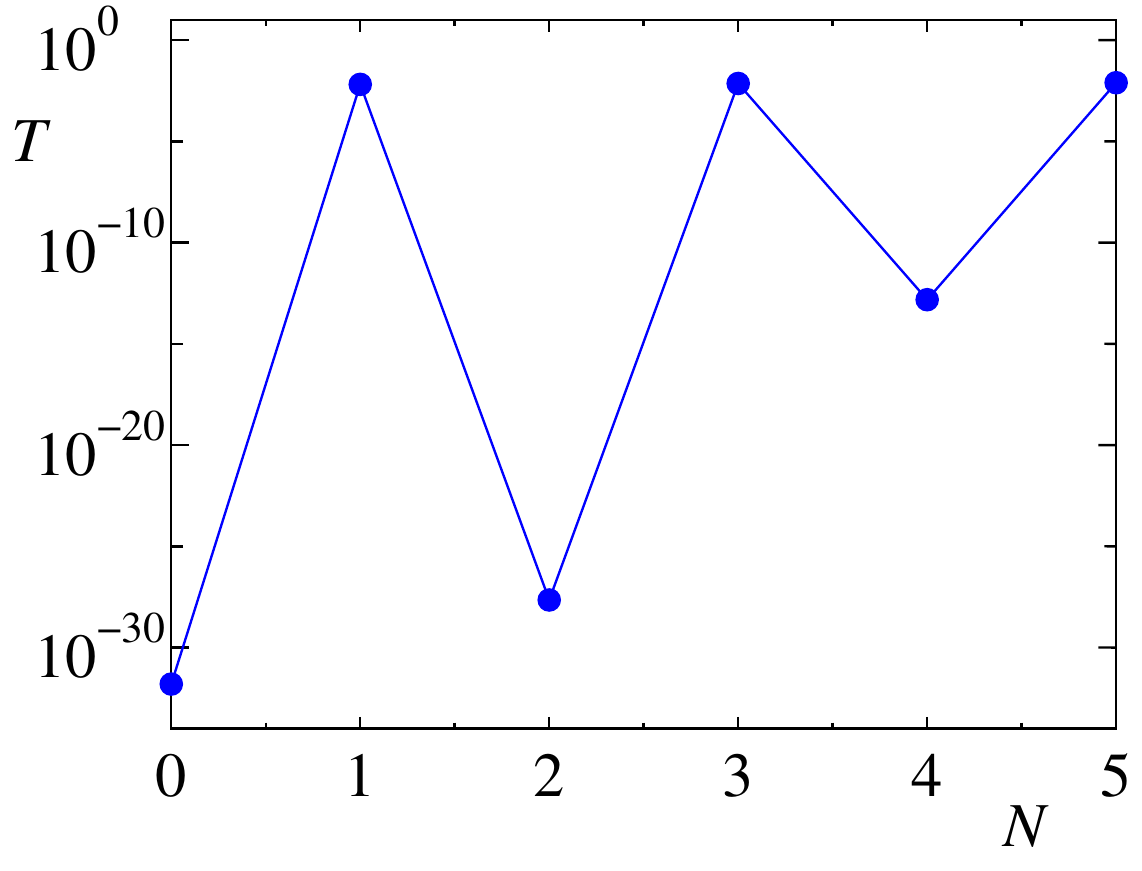}
\end{center}
\caption{
(Color online)
Numerical results of $T$ at $E_{\rm F}/m_{\parallel} = 0.0$
for $N = 0$, $1$, $2$, $3$, $4$, and $5$.
Although the values of $T$ for $N = 1$, $3$, and $5$
may look identical, they are slightly different from each other.
The solid line serves as a visual guide.
}
\end{figure}

Next, we consider the case with the side surface
under the open boundary condition in the $x$- and $y$-directions
to elucidate the effect of the helical surface states
as conduction channels connecting two electrodes.
Figure~4 shows the numerical results of $T$ in the presence and absence of
the screw dislocations for various values of $E_{\rm F}$
in the range of $0 \le E_{\rm F}/m_{\parallel} \le 2$.
We observe that the values of $T$ in the presence and absence of
the screw dislocations are of the same order of magnitude,
indicating that the contribution of the dislocation-induced helical modes
to $T$ is the same order of magnitude as, or smaller than,
that of the helical surface states.
This is because only one pair of screw dislocations is present in our setup.
Let us consider a situation in which many pairs of screw dislocations
are present in the system.
If the areal density of the dislocation pairs is kept constant,
the contribution of the dislocation-induced helical modes to $T$ is
proportional to $N_{x} \times N_{y}$, whereas that of
the helical surface states is roughly proportional to $N_{x} + N_{y}$.
Therefore, the contribution of the dislocation-induced helical modes to $T$
is dominated by that of the helical surface states
when both $N_{x}$ and $N_{y}$ are very small as in our setup.
Conversely, the contribution of the dislocation-induced helical modes
becomes much larger than that of the helical surface states
if both $N_{x}$ and $N_{y}$ are increased
while keeping the areal density of the dislocation pairs constant.
\begin{figure}[tbp]
\begin{center}
\includegraphics[height=4.0cm]{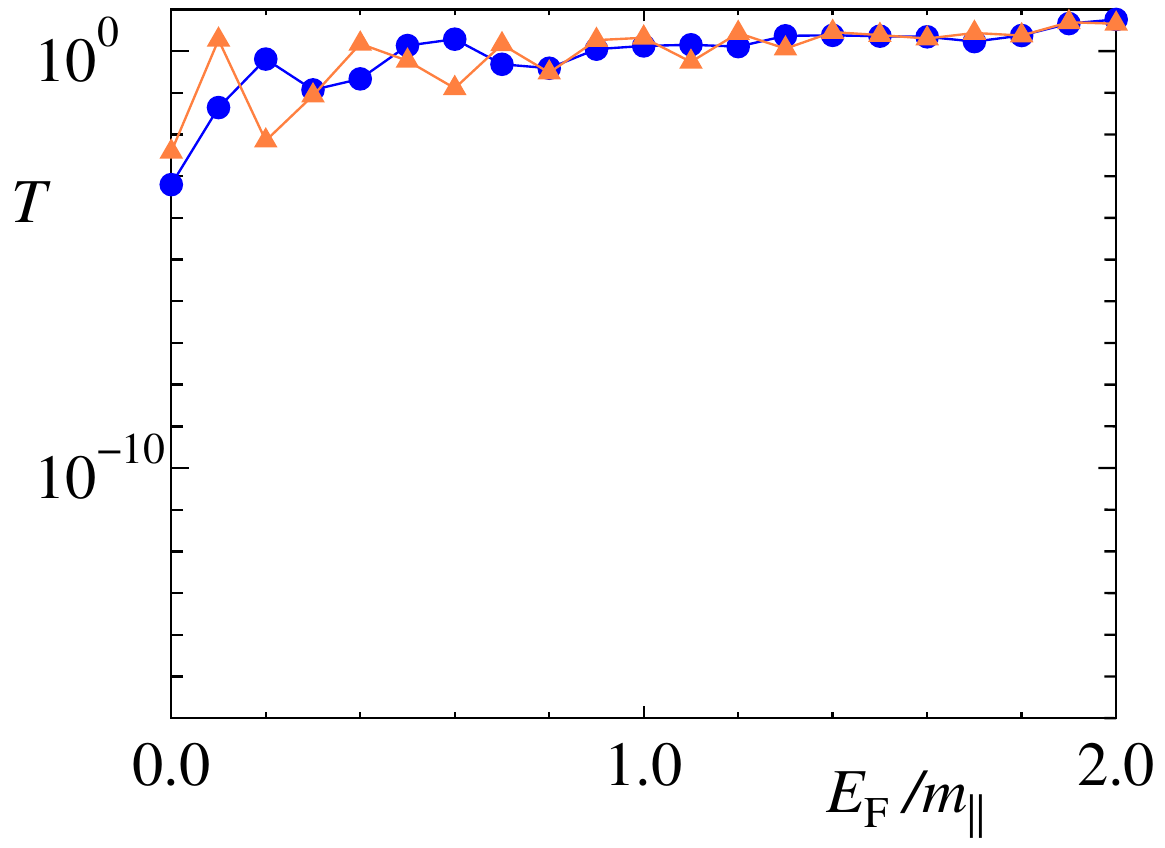}
\end{center}
\caption{
(Color online)
Numerical results of $T$ in the presence of the screw dislocations with $N = 1$
(triangles) and in the absence of the screw dislocations (circles)
for several values of $E_{\rm F}/m_{\parallel}$
under the open boundary condition in the $x$- and $y$-directions.
The solid lines serve as visual guides.
}
\end{figure}
\begin{figure}[btp]
\begin{center}
\includegraphics[height=2.5cm]{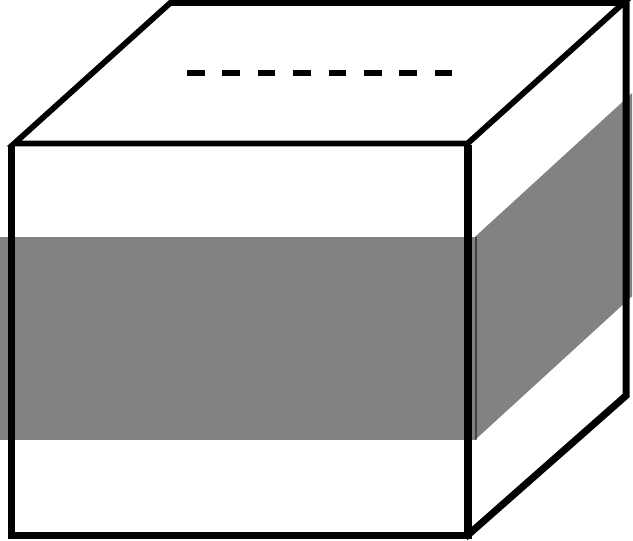}
\end{center}
\caption{
Schematic of the system with its side surface partly covered
with an insulating ferromagnetic layer.
The dashed line on the top surface represents the step edge
connecting two dislocation lines.
}
\end{figure}

Here, we briefly comment on the fluctuation in $T$
with the change in $E_{\rm F}/m_{\parallel}$.
This is mainly ascribed to finite size effects, which result in
the quantization of eigenstates.
Indeed, the fluctuation is large near $E_{\rm F}/m_{\parallel} = 0$
where quantization is particularly notable.

\begin{figure}[btp]
\begin{center}
\includegraphics[height=4.0cm]{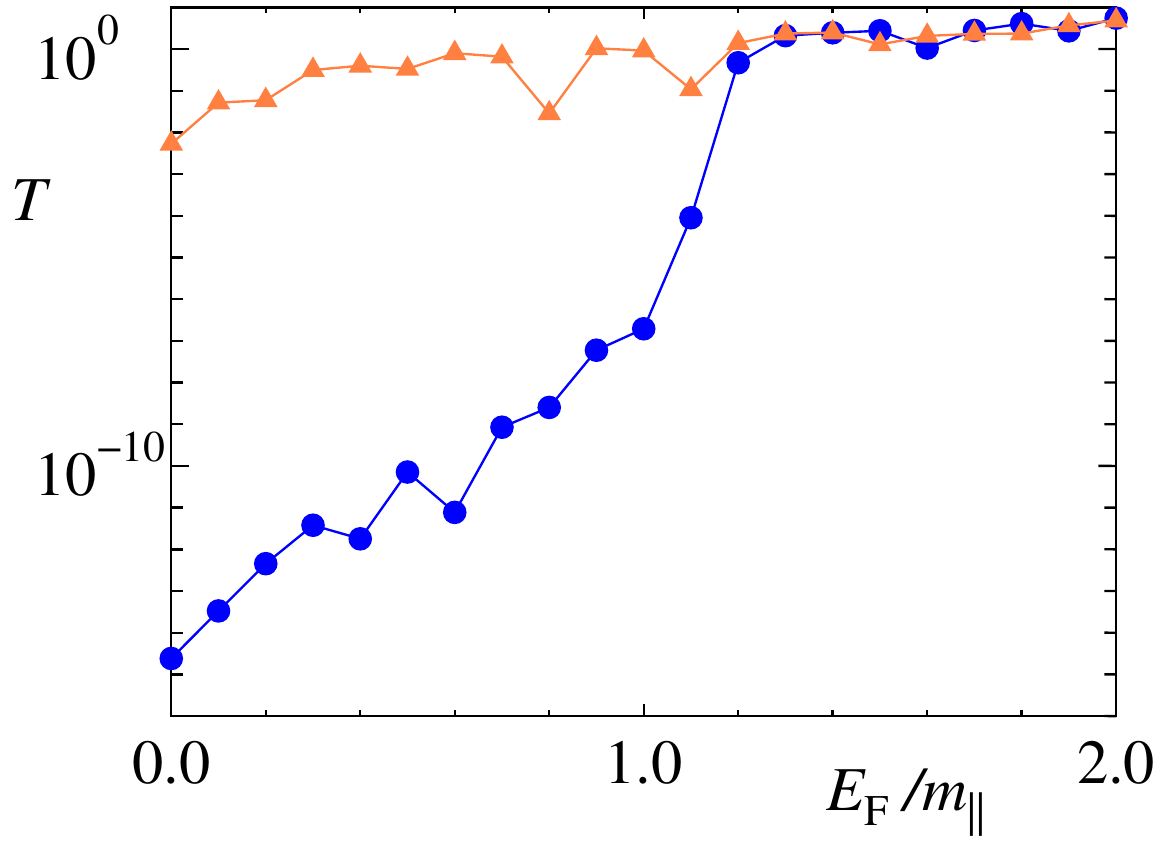}
\end{center}
\caption{
(Color online)
Numerical results of $T$ in the presence of the screw dislocations with $N = 1$
(triangles) and in the absence of the screw dislocations (circles)
when $m_{\rm mag}/m_{\parallel} = 1.2$
for several values of $E_{\rm F}/m_{\parallel}$
under the open boundary condition in the $x$- and $y$-directions.
The solid lines serve as visual guides.
}
\end{figure}
Finally, we show that, even in the case with the side surface
with small $N_{x}$ and $N_{y}$,
the dislocation-induced helical modes can dominantly contribute to $T$.
This is realized if an energy gap is opened in the spectrum of
the helical surface states on the side surface, resulting in
the blockade of the conduction channels through the helical surface states.
We introduce the energy gap by attaching surface magnetization to every site
on the side surface in the region of $n_{1}+1 \le n \le n_{2}$,
where $n_{1} = N_{z}/4$ and $n_{2} = 3N_{z}/4$.
This can be implemented by surrounding the sample with
an insulating ferromagnetic layer (see Fig.~5) such that
its magnetization is everywhere normal to the surface.~\cite{nomura1,nomura2}
The perturbation is expressed as
\begin{align}
   H_{\rm mag}
  = {\sum_{l, m}}'\sum_{n=n_{1}+1}^{n_{2}}
     |l,m,n \rangle \mathcal{H}_{\rm mag} \langle l,m,n|
\end{align}
with
\begin{align}
   \mathcal{H}_{\rm mag}
   = m_{\rm mag} \sigma_{\perp}(l,m) \otimes \tau_{0} ,
\end{align}
where $\tau_{0}$ is a $2 \times 2$ unit matrix
and the summation over $l$ and $m$ is restricted to
sites in the outermost layer of the side surface.
Here, $\sigma_{\perp}(l,m)$ is given as
\begin{align}
  \sigma_{\perp}(l,m)
  = \left\{
      \begin{array}{rl}
         -\sigma_{x} & (l = 1, \; 2 \le m \le N_{y}-1) , \\
          \sigma_{x} & (l = N_{x}, \; 2 \le m \le N_{y}-1) , \\
         -\sigma_{y} & (2 \le l \le N_{x}-1, \; m = 1) ,  \\
          \sigma_{y} & (2 \le l \le N_{x}-1, \; m = N_{y}) \\
      \end{array}
    \right.
\end{align}
except in the sites on the four corners, where it is given as
\begin{align}
  \sigma_{\perp}(l,m)
  = \left\{
      \begin{array}{rl}
          \frac{1}{\sqrt{2}}(-\sigma_{x}-\sigma_{y}) &
                     (l=1, \; m = 1) , \\
          \frac{1}{\sqrt{2}}(-\sigma_{x}+\sigma_{y}) &
                     (l=1, \; m = N_{y}) , \\
          \frac{1}{\sqrt{2}}(\sigma_{x}-\sigma_{y}) &
                     (l=N_{x}, \; m = 1) , \\
          \frac{1}{\sqrt{2}}(\sigma_{x}+\sigma_{y}) &
                     (l=N_{x}, \; m = N_{y}) .
      \end{array}
    \right.
\end{align}
Figure~6 shows the numerical results of $T$ in the presence and absence of
the screw dislocations when $m_{\rm mag}/m_{\parallel} = 1.2$
for various values of $E_{\rm F}$
in the range of $0 \le E_{\rm F}/m_{\parallel} \le 2$.
We observe that $T$ in the presence of the screw dislocations is
orders of magnitude larger than that in the absence of the screw dislocations
when $E_{\rm F}/m_{\parallel}$ is small.
By comparing the results when $m_{\rm mag}/m_{\parallel} = 0$ (Fig.~4)
and those when $m_{\rm mag}/m_{\parallel} = 1.2$ (Fig.~6),
we found that $T$ in the presence of the screw dislocations is
almost unaffected by $m_{\rm mag}$.
In contrast, $T$ in the absence of the screw dislocations strongly
reduces when $m_{\rm mag}/m_{\parallel} = 1.2$.
The reduction in $T$ in the absence of the screw dislocations can be ascribed
to the blockade of the conduction channels through the helical surface states
due to the energy gap induced by surface magnetization.~\cite{comment2}
The insensitivity of $T$ to $m_{\rm mag}$
in the presence of the screw dislocations
indicates that the conduction channels along the screw dislocations
are unaffected by surface magnetization.
The results show that the contribution of
the dislocation-induced helical modes to $T$
becomes more dominant than that of the helical surface states
if an energy gap is opened in the spectrum of
the helical surface states on the side surface.

\section{Summary and Short Discussion}

If screw dislocations are inserted into a strong topological insulator,
one-dimensional helical modes appear along each screw dislocation
under a certain condition.
To determine whether the dislocation-induced helical modes act as
one-dimensional conduction channels,
we calculate the two-terminal conductance $\mathcal{G}$ between a pair of
electrodes placed on the top and bottom surfaces
of a strong topological insulator, into which a pair of screw dislocations
is inserted such that it connects the top and bottom surfaces.
From the results of $\mathcal{G}$, we conclude that
the dislocation-induced helical modes certainly act as
one-dimensional conduction channels connecting the top and bottom surfaces.

Our conclusion is consistent with the results of
experimental studies,~\cite{hamasaki1,hamasaki2} in which
the conductivity in the presence of high-density dislocations
is enhanced in the direction parallel to the dislocations.
Unfortunately, a detailed comparison between our numerical results
and the experimental results is difficult.
The main reason for this is that our model system
with only one pair of screw dislocations is considerably smaller than
the experimental systems with high-density dislocations.
Another reason is that our tight-binding model cannot describe
a realistic feature of the experimental systems.
Indeed, in contrast to the fact that
the contribution of helical surface states to the conductivity
is less dominant than that of bulk states
in the experiment reported in Ref.~\citen{hamasaki1},
the former is much more dominant than the latter in our model
as long as the Fermi energy is in the subgap region.

\section*{Acknowledgment}

This work was supported by JSPS KAKENHI Grant Number JP21K03405.


\begin{thebibliography}{99}
\bibitem{comment1} The estimated energy gap is smaller than
the bulk energy gap $E_{\rm g}$ (see Sect.~2).
This is ascribed to the gap edge smearing due to finite size effects.
\bibitem{comment2} Roughly speaking, the energy gap induced in the spectrum of
the helical surface states is nearly equal to $m_{\rm mag}$,
which we set $m_{\rm mag}/m_{\parallel} = 1.2$.
This is consistent with the behavior of $T$ shown in Fig.~6.
Indeed, the quantitative difference between the values of $T$
in the presence and absence of the screw dislocations
almost vanishes when $E_{\rm F}/m_{\parallel} \gtrsim 1.2$.
\end{thebibliography}
\end{document}